# Vortex matter and strong pinning in underdoped PrFeAs(O,F) with atomic-sized defects.

Andrey V. Sadakov[1,*,†], Vladimir A. Vlasenko[1,†], A.Yu. Levakhova [1,†], I.V. Zhuvagin[1], E.M. Fomina[1], V.A. Prudkoglyad[1], A.Y. Tsvetkov[1], A.S. Usoltsev[1], and N. D. Zhigadlo[2,3]

1 P. N. Lebedev Physical Institute, Russian Academy of Sciences, Moscow 119991, Russia

2 CrystMat Company, CH-8037 Zurich, Switzerland

3 Laboratory for Solid State Physics, ETH Zurich, CH-8093 Zurich, Switzerland

**ABSTRACT**. We present a comprehensive investigation of the field-dependent critical current density and pinning force, combined with a detailed analysis of the nanostructural defect landscape in single crystal of underdoped PrFeAs(O,F) superconductor. Our study demonstrates that for both in-plane and out-of-plane magnetic field orientations critical current density exhibits a strong pinning regime in intermediate fields across the entire temperature range. The dominant contribution to pinning originates from oxygen-to-fluorine substitutional defects, oxygen vacancies, which all act as point defects via a quasiparticle mean free path fluctuation mechanism. Scanning transmission electron microscope studies did not reveal any volume or surface defect types within the lattice.

## I. INTRODUCTION.

The past decade has witnessed remarkable advancements in the field of superconductivity, marked by ground-breaking discoveries and transformative insights. A new class of high-temperature superconductors (HTSC), hydrogen-based hydrides, has emerged, exhibiting unprecedented superconducting properties under extreme pressures [1-5]. Furthermore, applied research on cuprate superconductors has seen pivotal breakthroughs, particularly in enhancing their performance and scalability for practical applications [6, 7]. Concurrently, significant progress has been achieved in understanding the mechanisms underlying iron-based superconductors (IBSC), providing insight into their complex electronic and magnetic interactions [8] superconducting order parameter structure [9, 10]. Iron-based superconductors are known to exhibit lower critical temperatures compared to most cuprate superconductors, however, they achieve upper critical fields and critical current densities that are competitive with those of cuprates [11]. Beyond this, the 1111 system has the highest critical temperature among other iron superconductors and critical current density up to $10^7$ A/cm$^2$ [12-14]. Moreover, this particular family shows weaker thermal fluctuations, compared to cuprates, which makes them highly attractive candidates for applications in high-field environments.

The performance and applicability of high-temperature superconducting (HTSC) materials in high and ultra-high magnetic fields are intrinsically linked to the properties and dynamics of the vortex phase. This phase emerges from a complex interplay of thermal and quantum fluctuations, intervortex interactions, and pinning potentials. The pinning potential, in particular, plays a pivotal role by immobilizing vortices, preventing energy dissipation. Therefore, the control and optimization of vortex pinning provide a powerful experimental lever to enhance critical parameters, such as the critical current density. Consequently, a comprehensive study of the mechanisms governing pinning forces in iron-based superconductors, coupled with detailed nanostructural characterization of the morphology and intrinsic nature of pinning centers, is critically essential for advancing their functional performance in practical applications.

To determine the primary contribution to flux pinning, the scaling of pinning force curves, $F_p=J_c*B$, is typically used, where $J_c$ is the critical current density and $B$ is the magnetic field. As shown in the works of Kramer [15] and Dew-Hughes [16], the nature of pinning centers can be inferred from the shape of the scaling dependence $f_p = F_p/F_p^{max}$ as a function of the reduced magnetic field $b = B/B_{irr}$ (where $B_{irr}$ is the irreversibility field). This approach generally works well for a variety of superconductors [17], however such a simple analysis is often insufficient for a thorough vortex pinning investigation and should be complemented by nanostructural studies of real defects landscape in the crystal. In this work, we present a comprehensive elemental and structural analysis of scanning electron microscopy images, in order to determine the most probable types of defects that contribute to pinning. Combined with detailed studies of irreversible magnetization loops, critical current densities, pinning force in both experimental geometries ($B//c$ and $B//ab$), we draw conclusions about the nature and characteristics of pinning in

PrFeAs(O,F) superconducting compounds. Finally, we build a complete phase diagram for both field directions.

## II. EXPERIMENTAL DETAILS

### A. Synthesis and characterization of PrFeAs(O,F) single crystals.

The PrFeAs(O,F) single crystals were synthesized via a cubic-anvil high-pressure high-temperature method. Starting materials comprising high-purity PrAs, $FeF_2$, $Fe_2O_3$, and Fe (≥99.95%) were precisely weighed in stoichiometric proportions, thoroughly ground in an agate mortar, and homogenized with NaAs flux. For each growth batch, 0.45 g of $PrFeAsO_{0.60}F_{0.35}$ and 0.2 g of NaAs were utilized. The synthesis involved heating the mixture to ∼1500°C over 2 hours, maintaining this temperature for 5 hours, followed by controlled cooling to 1250°C over 60 hours. The temperature was stabilized at 1250°C for 3 hours before cooling to ambient conditions. Crystalline products were isolated via dissolution of the flux matrix in distilled water. Additional synthesis specifics are documented in Ref. [18,19]. Most of the crystals displayed platelike shape with flat surfaces. The x-ray analysis confirmed that the obtained crystals belong to the 1111-type structure. Compositional analysis via energy dispersive x-ray (EDX) measurements confirmed that the ratio of praseodymium, iron, and arsenic is close to 1:1:1. Oxygen and fluorine cannot be measured accurately by EDX, therefore, we could not determine the exact doping level of the PrFeAs(O,F) crystals. Nevertheless, by a comparison of our transition temperatures with those of polycrystalline samples [20], one can estimate an F doping of 0.1, i.e., our crystals are underdoped.

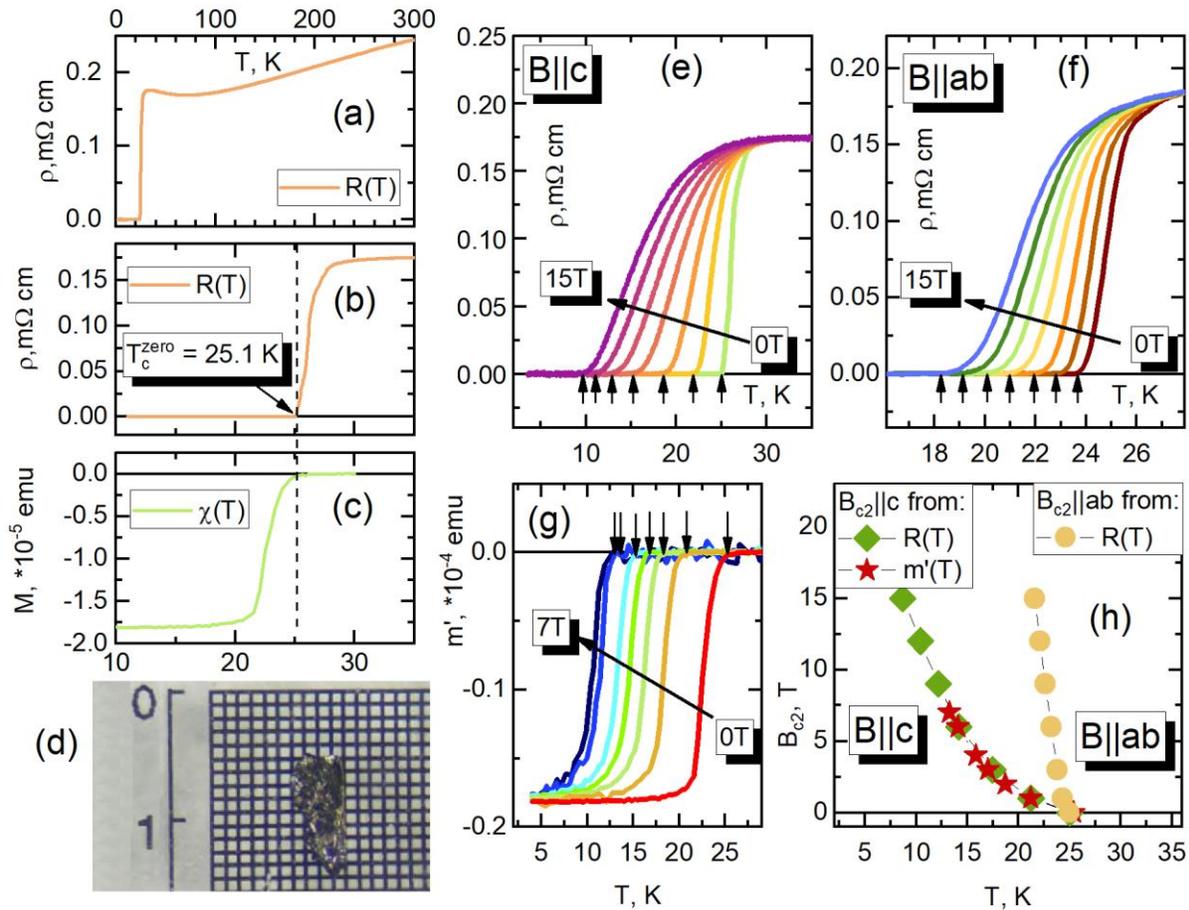

FIG 1. Transport and magnetic properties. (a) Temperature dependence of the sample resistivity R(T). (b), (c) Resistive transition and diamagnetic response of sample PrFeAs(O,F) (sample photograph in (d)), demonstrating the superconducting transition. The critical temperature $T_c$ is defined by zero resistivity and the onset of diamagnetic shielding. (e), (f) Field-dependent resistive transitions up to 15 T for B||c and B||ab, respectively. Black arrows mark the $B_{c2}$ determination criteria. (g) Superconducting transitions in magnetic susceptibility up to 7 T for B||c. (h) Temperature dependence of the upper critical field $B_{c2}$ for both field orientations.

**B. Magnetic and transport measurements.**

Magnetization measurements in high magnetic fields were performed using the MPMS-XL7, PPMS-9 by Quantum Design and 21T system manufactured by Cryogenic Inc., employing the vibrating sample magnetometry (VSM) method. The nominal sensitivity of the VSM magnetometer is $10^{-5}$ emu. The sensitivity of Squid is $10^{-7}$ emu. For the measurements, characteristic of the Meissner state to a nonlinear field-dependent magnetization response. This transition signals the beginning of bulk vortex penetration and subsequent establishment of critical flux density.

Transport measurements were carried out in a Cryogenic CFMS system, using a standart 4-prode method with a Kethley 6221 as a current source and a Keithley 2182a as a nanovoltmeter.

**C. STEM studies.**

For the transmission electron microscopy (TEM)

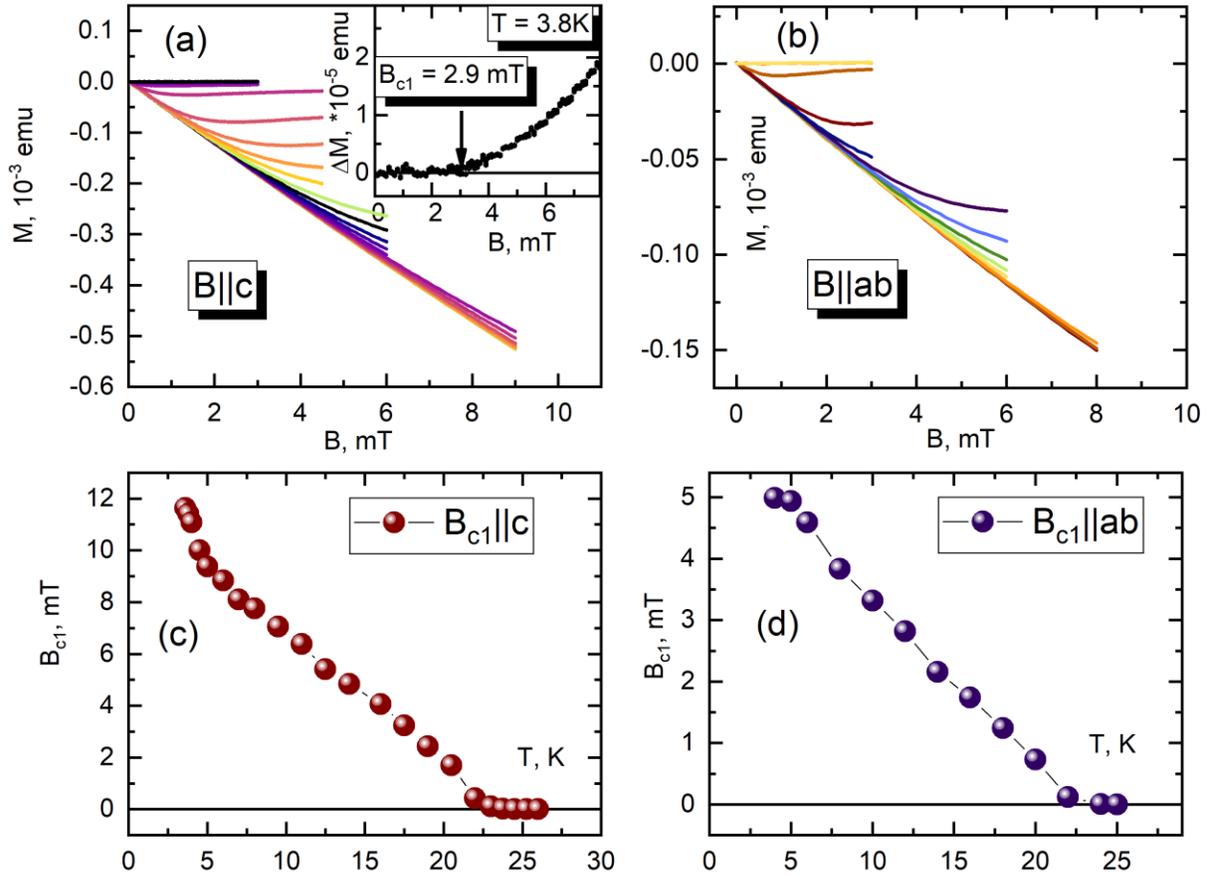

*FIG 2. (a) and (b) Magnetization curves of the PrFeAs(O,F) sample for B||c and B||ab orientations. The inset shows the criterion for determining the first critical field, $B_{c1}$. (c), (d) Temperature dependences of $B_{c1}(T)$ for both magnetic field orientations.*

the samples were mounted in a plastic non-magnetic holder. The experiments were conducted at fixed temperatures.

For determining the lower critical field ($B_{c1}$) we used the conventional approach, which relies on identifying the deviation from the linear $M(B)$ behavior study a cross-section of a single crystal was fabricated using a Helios G4 PFIB UXe dual-beam system (Thermo Fisher Scientific, MA, USA) using a focused ion beam (FIB) Xe+ plasma. A protective Pt layer of about 1.5 μm in total thickness was pre-deposited on the region of interest. The blank with dimensions of 6×6×2 μm was cut from the bulk crystal, fixed onto a copy moon-shaped FIB grid using an embedded EasyLift micromanipulator (Oxford Instruments, Abingdon, Oxfordshire, UK) and thinned at a low

**III. RESULTS AND DISCUSSION**

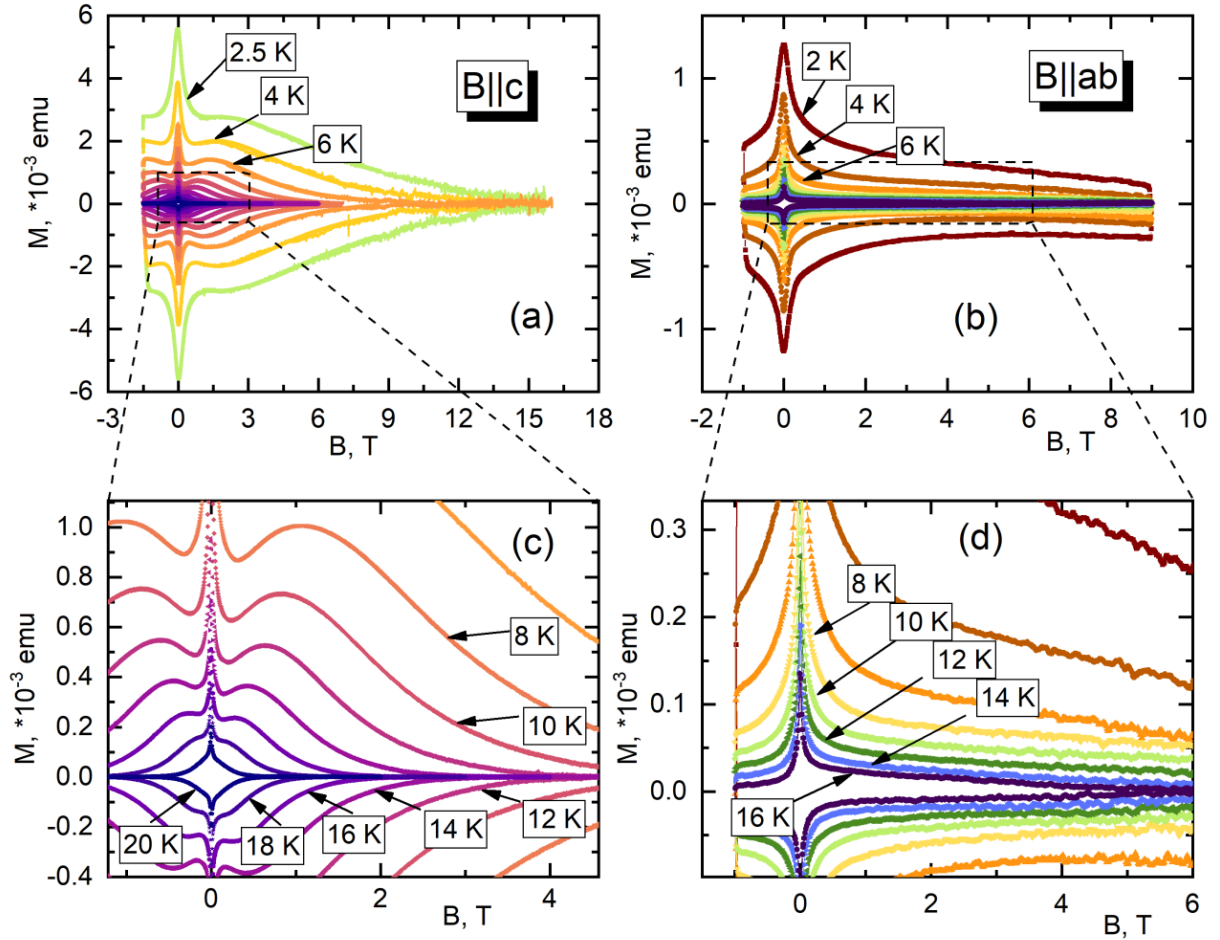

FIG 3. (a), (b) The isothermal magnetization loops, M(B), for the single crystal in two field orientations. (c), (d) Enhanced area of the magnetization data for higher temperatures.

beam current to a thickness of about 48 nm. The FIB grid with lamella was fixed in a double-tilt holder and placed in a TEM column. Electron diffraction (ED), high-angle annular dark field (HAADF-) and annular bright-field scanning TEM (ABF-STEM) images were acquired on a Titan Themis Z transmission electron microscope (Thermo Fisher Scientific, MA, USA) equipped with a DCOR+ condenser spherical aberration corrector, operated at 200 kV with a probe size of 0.3Å in STEM mode. The chemical composition of lamella was determined using energy-dispersive X-ray analysis (EDX) using the Super-X system, consisting of four large-area ring-shaped SSD detectors.

**A. Sample characterization and superconducting properties. Upper and Lower critical fields.**

The temperature dependence of the magnetic susceptibility and resistivity for the single crystal (PrFeAs(O,F) sample 1 photo is shown in Figure 1(d)) is presented in Figures 1(a-c). The critical temperature, is determined to be approximately $T_c^{zero}$ = 25.1 K, corresponding to the point where the resistivity vanishes and a pronounced diamagnetic response becomes evident. Temperature dependence of resistance measurements in magnetic fields up to 15 Tesla are shown in Figure 1(c) for the field orientation $B//c$ and in Figure 1(d) for $B//ab$. It is evident that in the presence of a magnetic field, the superconducting transition broadens and shifts toward lower temperatures. The broadening of the transition in the magnetic field is evidence that thermal fluctuations play a significant role in this compound. As is well known [21], the contribution of thermal fluctuations can be estimated from the value of the Ginzburg number, *Gi*:

$$Gi = \frac{1}{2}\left(\frac{k_B T_c}{4\pi\mu_0 B_{c2}^2(0)\epsilon\xi^3(0)}\right)^2 \quad (1),$$

where κ=λ/ξ is the Ginzburg-Landau parameter, $B_{c2}(0)$ is zero temperature upper critical field, $\epsilon$ is the anisotropy parameter, defined as $\epsilon = \xi_c/\xi_{ab}$.

The criterion of zero resistance was used to determine the upper critical field. Figure 1(g) shows the upper critical field dependence for both magnetic field orientations. For the B//c orientation, as a verification, the values of $B_{c2}(T)$ obtained from magnetic measurements (Figure 1(f)) in magnetic fields up to 7 Tesla are plotted. To estimate the $B_{c2}(0)$ values in both field directions, we used the WHH model [22, 23], in accordance with studies [24] on related systems of the 1111 class. The details of the WHH approximation and the corresponding curves are provided in the Supplemental Materials [25]. The obtained values of the upper critical field are as follows: $B_{c2}^{//c}(0) \approx 31T$ and $B_{c2}^{//ab}(0) \approx 90T$.

To obtain the temperature dependence of the lower critical field of the sample, we conducted magnetization measurements in both geometries. The magnetization versus magnetic field curves are presented in Figures 2(a) and 2(b). It is evident that the magnetization curves exhibit a distinct linear region corresponding to the Meissner state, in which Abrikosov vortices have not yet begun to penetrate the bulk of the superconductor. The lower critical field is determined as the field at which the magnetization curves deviate from the linear behavior. The criterion for $B_{c1}$ is marked by black arrow in the inset to Figure 1(a). The resulting temperature dependences of the lower critical field are displayed in Figures 1(c) and 1(d). The obtained values lower critical fields are as follows (taking into account demagnetization factor, obtained with accordance to ref. [26]): $B_{c1}^{//c}(0) \approx 12$ mT, $B_{c1}^{//ab}(0) \approx 5$ mT.

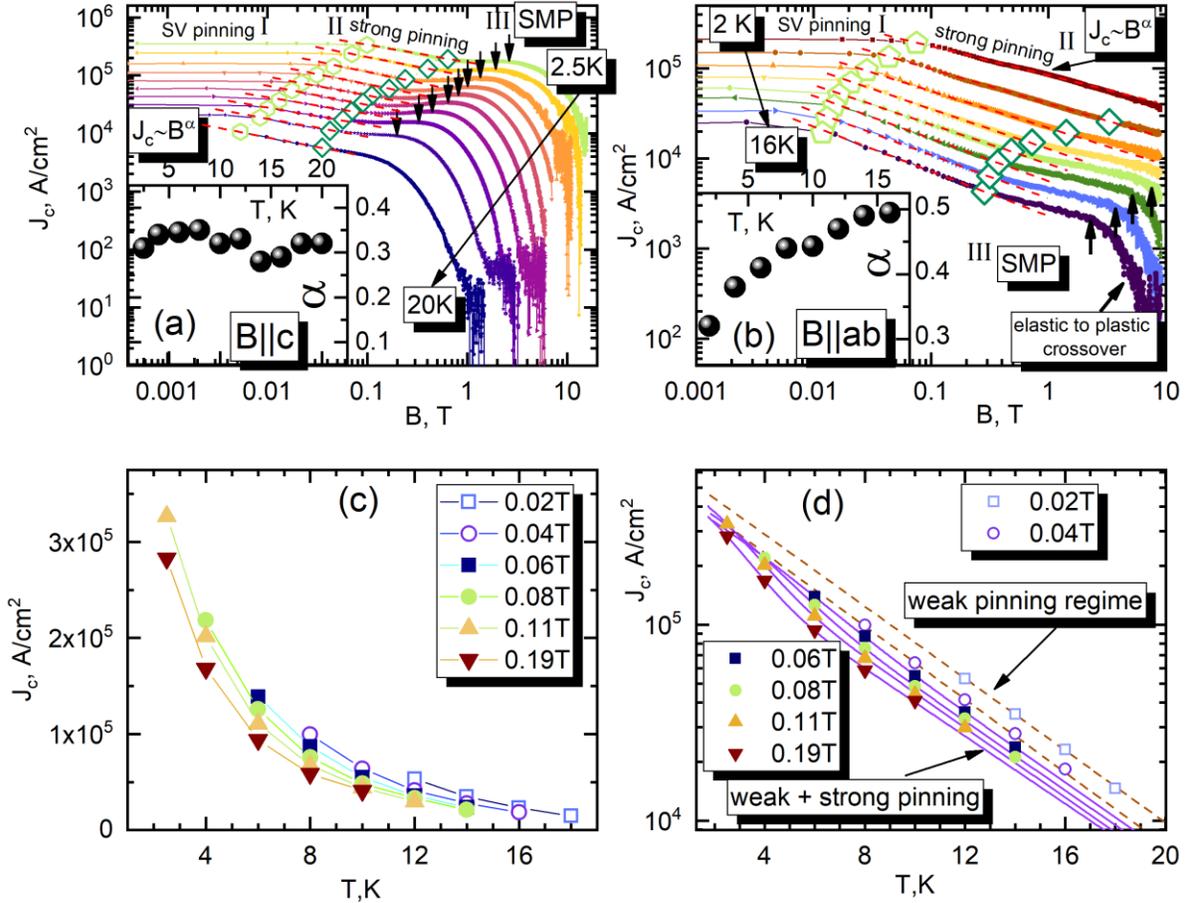

FIG 4. (a), (b) The $J_c(B)$ dependence in a double logarithmic scale for two magnetic field orientations. Highlighted regions include the (I) single-vortex pinning (SV) regime, (II) the power-law decay $J_c \sim B^{-\alpha}$ (dashed lines indicate linear fits), and (III) the second magnetization peak (denoted by black arrows). Insets: T-dependence of the power-law exponent, α. (c) The temperature dependence of the critical current density $J_c(T)$, derived from isofield sections of the $J_c(B)$ curves in (a). (d) fitting of the $J_c(T)$ data with weak and strong pinning models.

From the obtained dataset on the lower and upper critical fields in both magnetic field orientations, we estimate the values of the coherence length and penetration depth. To extract the anisotropic coherence length values, we used the formulas: $\xi_{ab} = \sqrt{\Phi_0/2\pi B_{c2}^c}$, and $\xi_c = \Phi_0/2\pi\xi_{ab}B_{c2}^{ab}$, where $\Phi_0$ is the flux quantum. For determining the London

$B_{c1}^c = (\Phi_0/4\pi\lambda_{ab}^2)ln\kappa_c$, and $B_{c1}^{ab} = (\Phi_0/4\pi\lambda_{ab}\lambda_c)ln\kappa_{ab}$, where $\kappa_c$ and $\kappa_{ab}$ are the Ginzburg-Landau parameters, obtained from formulas $\kappa_c = \lambda_{ab}/\xi_{ab}$ and $\kappa_{ab} = \sqrt{\lambda_{ab}\lambda_c/\xi_{ab}\xi_c}$. The calculated values are presented in table I.

TABLE I. Thermodynamic parameters describing the superconducting state of the PrFeAs(O,F) single crystal.

| $\xi_c$,nm | $\xi_{ab}$,nm | $\lambda_c$,nm | $\lambda_{ab}$,nm | $\kappa_c$ | $\kappa_{ab}$ |
|---|---|---|---|---|---|
| 1.14 | 3.23 | 901 | 251 | 78 | 247 |

### B. Magnetic hysteresis loops, critical current densities and pinning force.

Figures 3(a)-3(b) show the magnetization hysteresis loops $M(B)$ measured at fixed temperatures between 2 and 20 K in applied magnetic fields up to 19 T for the

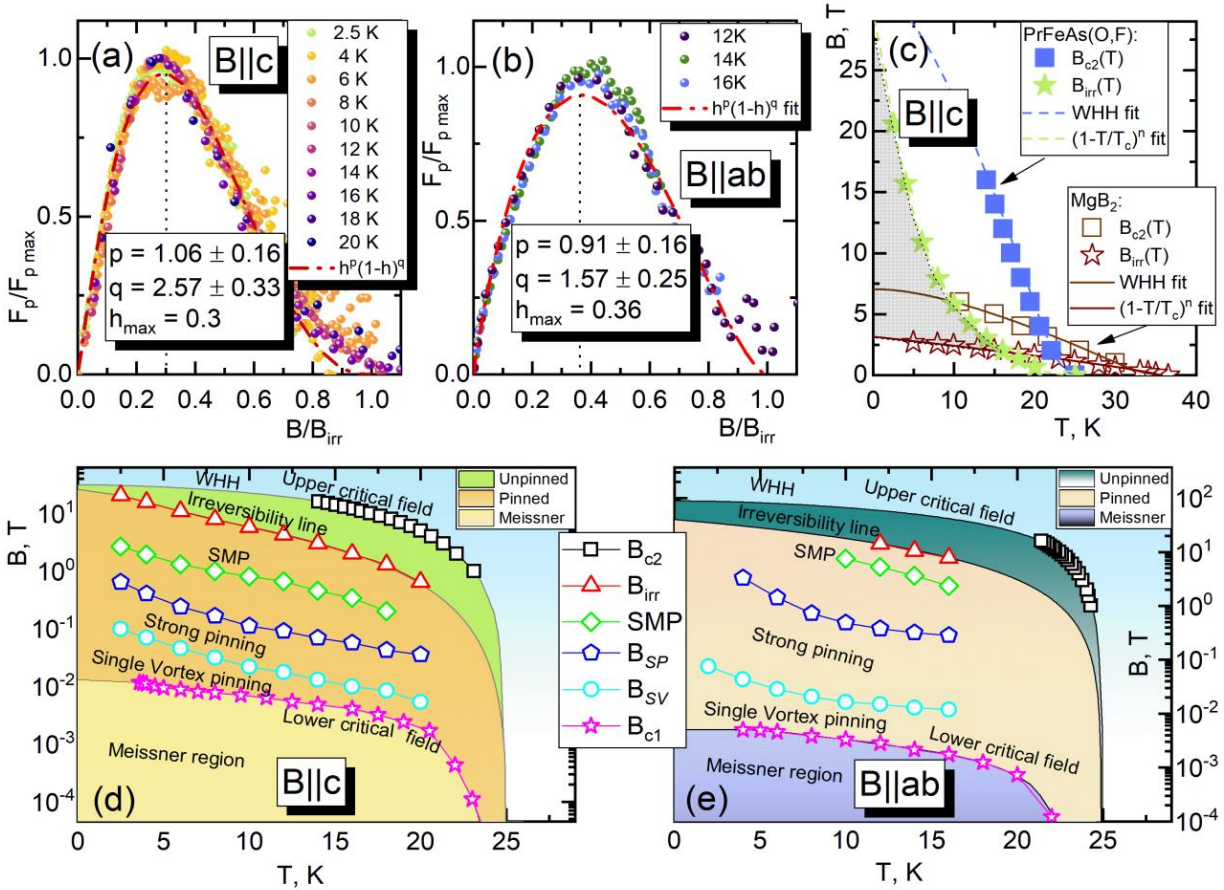

FIG 5. (a) and (b) Scaling dependencies of the reduced pinning force on the reduced magnetic field for the magnetic field orientations B//c and B//ab. The dash-dotted line represents the approximation using the Dew-Hughes model. (c) Comparison of the magnetic phase diagrams of a PrFeAs(O,F) single crystal (filled symbols for experimental data and dashed lines for theoretical fits) and an MgB$_2$ single crystal (hollow symbols for experimental data and solid lines for fits). The gray area indicates the region of the phase diagram where PrFeAs(O,F) is still capable of carrying supercurrent, while MgB$_2$ is not. (d) and (e) Magnetic phase diagram of PrFeAs(O,F) for both magnetic field directions. Symbols indicate experimental data points, while solid lines represent theoretical models.

penetration depth, we used the following expressions:

single crystal in both principal orientations (B//ab and

$B//c$). The hysteresis loops decrease in magnitude with increasing temperature while maintaining symmetric shapes, indicating that bulk pinning predominates over surface or geometric barrier effects.

Figs. 3(c)-3(d) zoom into the high-temperature area of the plot, showing the details and evolution of the hysteresis loops for both directions. A well-defined second magnetization peak (SMP) appears for $B//c$ below 20 K, as can be seen in Figs. 3(a) and 3(c), with the peak position shifting to higher fields upon cooling. In contrast, no SMP is observed for $B//ab$ within the measured field range (up to 9 T for this orientation), implying either its significant suppression in the *ab*-plane geometry or that its position is shifted to much higher fields.

The field dependence of the in-plane critical current density $J_c^{ab}$ for $B//c$ was determined using the standard Bean model [27], with the assumption of isotropic interplane currents ($J_c^a = J_c^b$). For $B\|ab$ geometry, the out-of-plane critical current density $J_c^c$ was determined via the anisotropic extension of the Bean model [28].

Figure 4 (a, b) (for the geometries $B//c$ and $B//ab$, respectively) presents the dependence of the critical current density on the magnetic field at fixed temperatures, plotted on a double logarithmic scale. It is evident that the curves exhibit similar behavior while displaying distinct regions with a well-defined $J_c(B)$ dependence. In the lowest fields, the critical current density is nearly independent of the magnetic field, corresponding to the single-vortex (SV) pinning regime. In the this regime, under weak magnetic fields, the intervortex distance, $a_0 = \sqrt{\Phi_0/B}$, is sufficiently large such that the critical current is governed solely by the pinning strength of individual vortices [21]. Thus, the critical current density remains independent (section I) of the magnetic field until vortices start interacting with each another. At intermediate fields (section II), a monotonic transition to a power-law dependence $J_c(B)$ is observed. The exponent in this dependence, which reflects the strength of pinning, was determined for each isothermal curve. Its temperature dependences are shown in the insets to Figure 4(a, b). For the $B\|c$ case, the exponent remains constant ($\alpha = 0.3$), whereas for the case where the magnetic field is aligned with the *ab* planes, the exponent gradually increases with temperature from 0.3 to 0.5. These values are characteristic of the strong pinning regime, as reported in [29, 30].

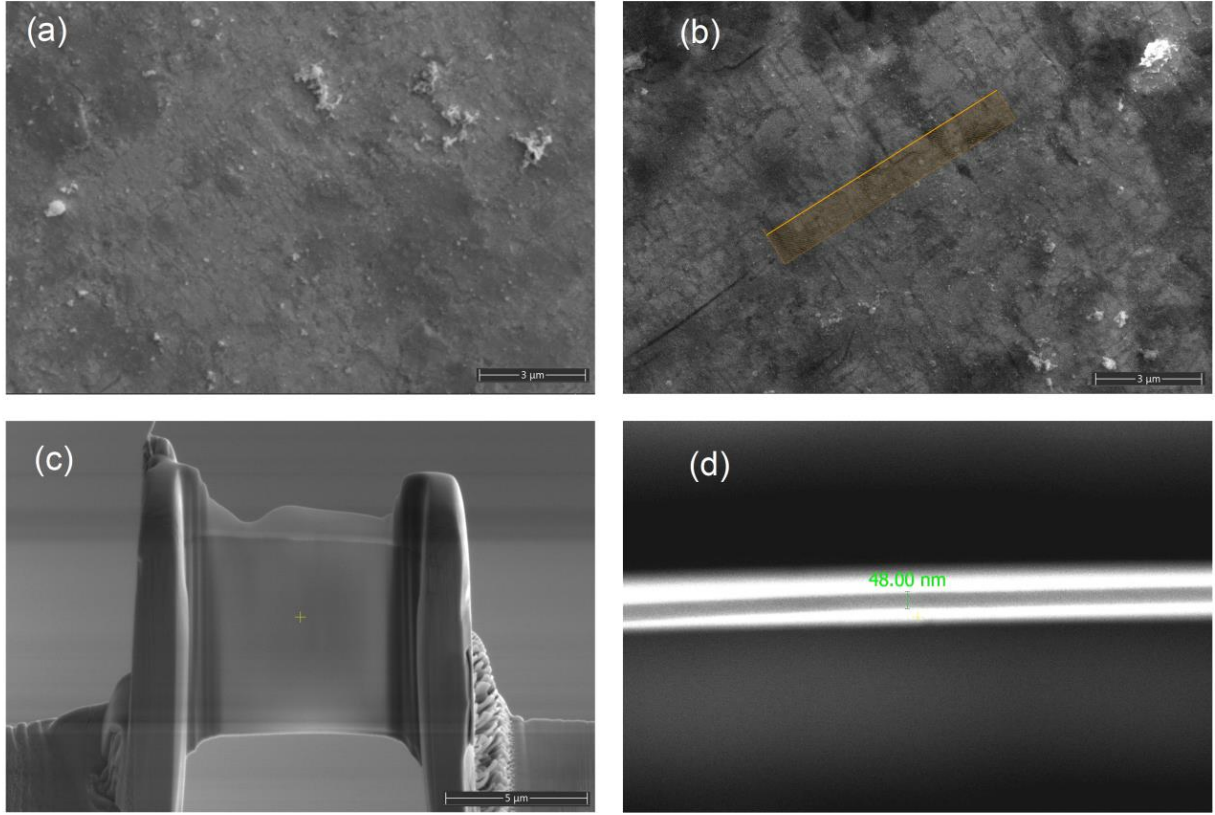

*FIG 6. (a) Electron microscope micrograph of the PrFeAs(O,F) crystal surface and the area from which the lamella was extracted. (c) Side view of the lamella (ac-plane). (d) Top view showing the thickness (along the b-axis) of the lamella.*

With further increases in the field (section III) for $B//c$, a plateau or slight increase in the critical current density is observed, corresponding to the second magnetization peak (SMP) that follows. For the longitudinal orientation, such a plateau emerges only at the highest temperatures, likely because the regime responsible for the SMP in this geometry is activated at significantly higher magnetic fields than those accessible in this experiment. At even higher fields, a sharp drop in $J_c(B)$ is observed, which is usually linked to an elastic-to-plastic transition in the vortex system [31-33]. The elastic regime describes a state where the vortex structure deforms reversibly under external perturbations, such as pinning forces or thermal fluctuations, while preserving its ordered, rigid structure. The rigidity of this vortex solid is primarily determined by the shear modulus $c_{66}$, which governs resistance to shear deformations, where one part of the lattice slides past another. According to ref. [34], the expression for $c_{66}$ is:

$$c_{66} \approx (B\Phi_0/16\pi \lambda_{ab}^2 \mu_0)(1 - B/B_{c2})^2 \quad (2).$$

Here, as per the expression (2), in low and intermediate magnetic fields, the shear modulus $c_{66}$ increases, maintaining the stiffness of the system. However, at a certain field, the shear modulus begins to decrease, softening the vortex structure. This softening allows vortices to occupy previously vacant pinning sites, resulting in an enhanced critical current density. This field marks the onset of the SMP. Finally, as the field approaches the upper critical field, shear modulus decreases significantly, and the vortex solid loses its ability to resist deformations elastically, giving place to an irreversible plastic vortex regime.

In order to clarify the pinning regimes active in our samples, we analyzed the $J_c(T)$ curves within a number of well-known pinning models. In ref. [21] it was theoretically established (see formula 2.75), that in the presence of weak pinning centres, the critical current density $J_c$ decays exponentially with increasing temperature, following the expression:

$$J_c^{weak}(T) = J_c^{weak}(0)e^{-T/T_0} \quad (3),$$

where $J_c^{weak}(0)$ is the zero temperature value of $J_c$, and $T_0$ is the characteristic energy of weak pinning centers. Experimentally these results were observed in cuprates and IBSCs in refs. [35, 36]. Subsequently, for HTSCs with columnar defect, which exhibit strong pinning, Nelson and Vinokur [37] and Hwa [38] predicted a temperature-dependent decay of $J_c$, described by the expression:

$$J_c^{strong}(T) = J_c^{strong}(0)e^{-3(T/T^*)^2} \quad (4),$$

where $J_c^{strong}(0)$ is the contribution of strong pinning centers to the $J_c$ at zero temperature and $T^*$ characterizes the vortex pinning by strong defect centers. Later experimental results demonstrated that this dependence is accurately captures the behavior of the critical current density in systems with strong pinning, including those that go beyond columnar defect types [36, 39]. Finally, studies in [36, 40] revealed that certain systems exhibit coexistence of both weak and strong pinning types, with the resulting $J_c(T)$ dependence approximated in its simplest form as:

$$J_c(T) = J_c^{weak}(0)e^{-T/T_0} + J_c^{strong}(0)e^{-3(T/T^*)^2} \quad (5).$$

Strictly speaking, the precise derivation of this dependence necessitates the summation of the pinning forces acting on a variety of interacting vortices, which is conceptually challenging task beyond the scope of this work.

The $J_c(T)$ dependencies derived from vertical constant field cuts of the $J_c(B)$ data are shown (figure 4(c)). At low magnetic fields ($B \leq 0.04$ T), the best fit was achieved using equation (3), presented by straight dashed lines in semilog-scale in figure 4(d). For intermediate fields ($0.06 < B < 0.19$ T), the $J_c(T)$ dependence can no longer be described by a simple exponential function. Moreover, formula (4) for strong pinning also fails to provide a satisfactory approximation. Therefore, we employed formula (5), which accounts for both contributions. The results of all approximations are represented by solid curves in figure 4(d). Thus, at a magnetic field of approximately 0.06 T, a crossover occurs from weak pinning to a combined pinning regime involving both weak and strong pinning contributions.

We proceed to analyze the field dependence of the pinning force and the scaling of the $f_p(b)$ curves in our crystals. The data were processed within the Dew-Hughes model [16] for both magnetic field orientations. To determine the reduced magnetic field $b=B/B_{irr}$, we employed the irreversibility field $B_{irr}$ derived via Kramer's method [15].

In this approach, the data are plotted in $J_c^{0.5}B^{0.25}$ vs $B$ coordinates, and $B_{irr}$ is defined as the intersection point of the linear extrapolation of the curve with the x-axis (see Supplemental Materials [25]). For the $B//c$ orientation, $B_{irr}$ values were successfully obtained across the entire temperature range (Figure 5(a)), while for the $B//ab$ orientation, only the highest temperatures yielded reliable $B_{irr}$ values (Figure 5(b)). Notably, both out-of-plane and in-plane orientations contributions from other types of pinning centres, such as grain boundaries, crystal surfaces, and others. The same measurements and analysis were performed for PrFeAs(O,F) sample 2, which exhibits a slightly higher critical temperature $T_c$. Corresponding magnetic hysteresis loops, field-dependent critical current density $J_c(B)$, and pinning force $F_p(B)$ are presented in the Supplemental Material [25]. The results obtained showed good agreement between the two samples.

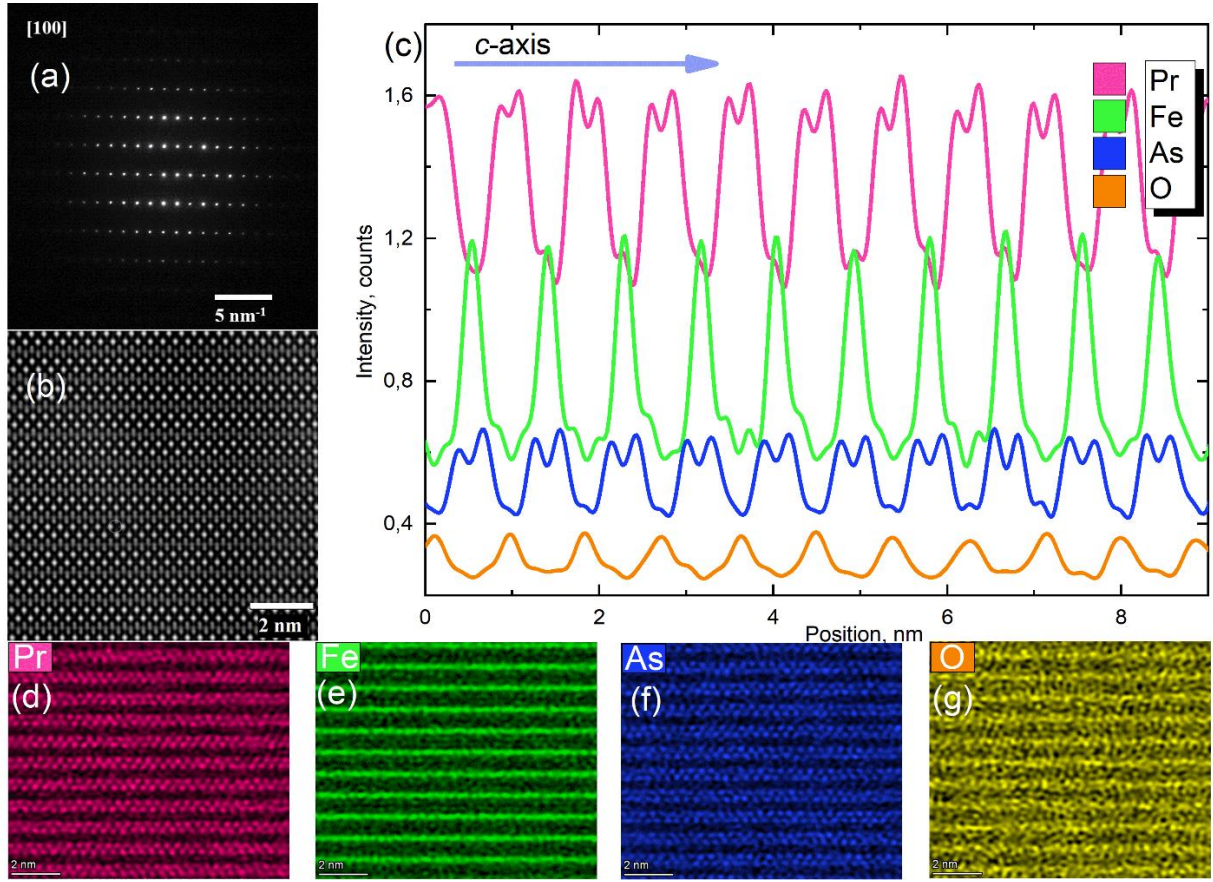

*FIG 7. (a) [100] SAED. (b) HR HAADF STEM image with atomic structure of PrFeAs(O,F). (c) Intensity profiles by element from (a) along the c-axis. (d–g) EDX elemental maps.*

exhibit excellent scaling of the $f_p(b)$ curves into unified dependencies. The results of the Dew-Hughes model fitting are also shown in Figures 5(a) and 5(b). The extracted scaling parameters $p$, $q$, and the peak positions are as follows: $p = 1.06$, $q = 2.57$, $h_{max} = 0.3$ for $B//c$, and $p = 0.91$, $q = 1.57$, $h_{max} = 0.36$ for $B//ab$. According to the Dew-Hughes theory, these parameters most closely align with the model of pinning by point defects: $p = 1$, $q = 2$, $h_{max} = 0.33$. Thus, the observed $f_p(b)$ scaling suggests that point defects likely dominate the pinning mechanism. The slight deviation from the tabular values in the approximation is likely attributed to additional pinning

Additionally, we conducted a comparison of the superconducting properties of PrFeAs(O,F) and another well-known superconductor, $MgB_2$. We selected magnesium diboride specifically because this compound is considered a highly promising superconducting material and has a critical temperature close to that of our superconductor. Figure 5(c) shows the phase diagrams of both superconductors, obtained from single-crystal samples. The data for magnesium diboride were taken from Refs. [41][42]. It can be seen that although PrFeAs(O,F) has a lower critical temperature, the upper critical field curve and, more importantly, the irreversibility line at low temperatures are significantly higher than those of magnesium diboride. This makes PrFeAs(O,F) potentially more attractive for practical applications, especially considering its lower anisotropy.

Finally, the complete magnetic phase diagrams of PrFeAs(O,F) were constructed for both field orientations across a magnetic field range from $10^{-4}$ to $10^2$ Tesla (Figures 5 (a), (b)). The diagrams delineate key phases: the Meissner state, single-vortex pinning regime, strong pinning regime, second magnetization peak, and unpinned vortex phase. It is noteworthy that, aside from a slight difference in the width of certain regions, the phase diagrams for the two orientations show a remarkable closeness.

### C. Microstructure study by high resolution STEM.

Figure 6 shows micrographs of the PrFeAs(O,F) single crystal surface and the region from which a lamella was cut for transmission electron microscopy studies. The lamella is a parallelepiped with an *ac*-plane dimensions of approximately 6×6 μm and a thickness of 48 nm (figures 6(c), (d)). The lamella was subsequently loaded into the STEM chamber. The selected area electron diffraction (SAED) in [100] direction is shown in Figure 7(a). Figure 7(b) presents HR HAADF STEM atomic structure of the PrFeAs(O,F) at 300 K. Neither bulk nor planar defects were detected in the sample (Figure 7b), unlike observations in the 111 and 1144 systems [43, 44], where low concentrations of planar defects associated with embedded 122-type secondary phases were observed. In situ elemental analysis via EDX method along the *c*-axis (intensity profile in vertical direction in Figure 7b) is shown in Figure 7c. Distinct intensity peaks corresponding to Pr, Fe, As, and O atoms are clearly visible. In Figure SM8, we present an enlarged section of the intensity profile plot overlaid with the crystal lattice. The lattice parameter $c = 0.872$ nm, obtained from these data, shows good agreement with values from X-ray diffraction analysis [19]. We were unable to resolve the fluorine intensity profiles for two reasons: first, light elements are poorly detected by the EDX technique, especially at such low concentrations; second, the fluorine K$\alpha_1$ line (676.8 eV) is too close to the iron L$\alpha_1$ line (705 eV), making their experimental distinction practically very difficult. Consequently, transmission electron microscopy (TEM) studies established the absence of any extended or bulk defect types (dislocations, grain boundaries, twin boundaries etc.) in the examined region of the sample. Thus, the only defect type most likely present in our sample is point defects, such as oxygen substitution by fluorine and oxygen vacancies, which could not be resolved by the available methodology.

### D. Discussion.

In order to precisely determine the nature and mechanism of pinning, we focus on generalizing the conclusions from Sections B and C. Analysis of the reduced pinning force's dependence on the magnetic field clearly indicates that point-like pinning centres of the $\delta l$ type dominate the pinning behavior in both magnetic field orientations. In this scenario, pinning arises from variations in the mean free path within the vortex core, where dopant atoms and vacancies serve as effective scattering centres for quasiparticles. According to van der Beek [45], in the single-vortex pinning regime, the critical current density is given by:

$$J_c^{SV} \approx J_0 \left[ \frac{0.1 n_d D_v^4}{\xi_{ab} \lambda_{ab/\lambda_c}} \left(\frac{\xi_0}{\xi_{ab}}\right)^2 \right]^{2/3}, \qquad (6)$$

here, $J_0 = 4\epsilon_0/\sqrt{3}\Phi_0 \xi_{ab}$ is the depairing current density, $n_d$ is the defect density, $D_v$ is the defect radius, $\epsilon_0 = \Phi_0^2/4\pi\mu_0 \lambda_{ab}^2$ is the typical vortex energy scale, $\xi_0 \cong 1.35\xi(0)$ is temperature independent BCS coherence length. If we assume that the dominant contribution to pinning arises from fluorine substitutional defects at oxygen sites (and, to a lesser extent, oxygen vacancies), then the defect size $D_v$ can be approximated as the average ionic radius of oxygen and fluorine, $D_v = 0.135\ nm$. In this case, the single vortex critical current density $J_c^{SV}$ at the lowest temperature, calculated via Eq. (6), can be achieved with a defect concentration of $n_d = 8.2 \times 10^{27}\ m^{-3}$ corresponding to approximately 0.59 defects per unit cell.

Furthermore, in our view, Figure 4 shows a rather unexpected result: the field dependencies of the critical current density for the two magnetic field orientations are remarkably similar, with only minor differences. This similarity is particularly surprising given the visually and quantitatively distinct magnetic moment hysteresis loops for each orientation. In fact, all key regimes in the $J_c(B,T)$ phase diagram are observed in both the longitudinal and transverse field configurations: the single-vortex pinning regime at low magnetic fields, the strong pinning regime at intermediate fields, and the second magnetization peak, which is accompanied by a crossover from elastic to plastic vortex behavior. The primary differences are twofold: a slight temperature dependence of the exponent in the strong pinning regime and a significant shift of the elastic-to-plastic (E-P) crossover to higher fields for the longitudinal orientation.

In our opinion, the most plausible explanation for this phenomenon is a combination of several factors. First, the fundamental parameters of our system, such as the coherence length and the penetration depth, exceed the distance between the superconducting Fe-As planes. This distinguishes our system from, for example, the cuprates [46] [47] and some other IBSC [48], where vortices or their segments can be expelled into the spacers between the superconducting planes, forming so-called pancake vortices or Josephson vortices. Second, our system contains a very low density of extended defects (dislocations, twin boundaries, grain boundaries) that could provide vortices with an easy direction for pinning at specific pinning centers. The primary pinning landscape in our case consists of an isotropic system of randomly distributed point defects. Therefore, we suggest that this is the reason the field dependencies of the critical current density appear so similar for the two orientations of the magnetic field.

## IV. CONCLUSIONS

In conclusion, this work presents a comprehensive investigation of the vortex physics and its interplay with the landscape of nanostructural defects in a PrFeAs(O,F) crystal. The behavior of the critical current density was systematically studied as a function of temperature, magnetic field strength, and field orientation with respect to the crystal planes. It was demonstrated that the critical current density exhibits remarkably similar trends for both magnetic field orientations, revealing distinct regimes: single-vortex pinning, strong collective pinning, a second magnetization peak associated with the onset of vortex elasticity, and a crossover from elastic to plastic deformations of the vortex lattice.

With the mean-free-path fluctuation mechanism, point defects provide the dominant contribution to the pinning. Quantitative analysis estimates the concentration of these defects to be 0.59 per unit cell. HRTEM investigations, on the other hand, revealed no extended pinning centers (bulk, surface, dislocations, etc.), leaving contributions from oxygen vacancies and O/F substitution as the most probable.

Finally, a detailed magnetic phase diagram of the vortex state in the superconducting PrFeAs(O,F) compound was constructed for both magnetic field orientations, providing a complete framework for understanding the interplay between defect microstructure and vortex dynamics.

## ACKNOWLEDGMENTS
We thank V. Geshkenbein and A. Shilov for helpful comments. A.V.S. acknowledges the financial support of RSF grant No. 23-12-00307.

## CONTRIBUTIONS
N.D.Z. synthesized the single crystals. A.V.S. coordinated all experimental efforts, processed and analyzed the magnetic data, and wrote the manuscript. V.A.V. contributed to data analysis and manuscript preparation. A.Yu.L. performed STEM measurements and participated in manuscript drafting. I.V.Zh. conducted transport measurements on single crystals in magnetic fields up to 15 T, and processed and analyzed the corresponding data. E.M.F. carried out magnetic measurements using a SQUID magnetometer and processed the data. V.A.P. performed high-field magnetic measurements on a 21T system. A.Y.T. conducted measurements using a PPMS-9 system. A.S.U. contributed to data processing and analysis.